\documentclass[12pt]{article}

\usepackage{epsf,epsfig,psfig}

\def\spose#1{\hbox to 0pt{#1\hss}}
\def\ltapprox{\mathrel{\spose{\lower 3pt\hbox{$\mathchar"218$}}
 \raise 2.0pt\hbox{$\mathchar"13C$}}}
\def\gtapprox{\mathrel{\spose{\lower 3pt\hbox{$\mathchar"218$}}
 \raise 2.0pt\hbox{$\mathchar"13E$}}}
\def\case#1#2{{\textstyle\frac{#1}{#2}}}

\begin{document}
\begin{titlepage}
\begin{center}
{\Large\bf The critical behavior of magnetic systems described by
Landau-Ginzburg-Wilson field theories}
\end{center}
\vskip 1.3cm
\centerline{
Pasquale Calabrese,$^a$ 
Andrea Pelissetto,$^b$ 
Ettore Vicari$^{c\;*}$}

\vskip 0.4cm
\centerline{\sl  $^a$ Scuola Normale Superiore, Pisa, Italy} 
\centerline{\sl  $^b$ Dipartimento di Fisica dell'Universit\`a di Roma I,
                       Roma, Italy}
\centerline{\sl  $^c$ Dipartimento di Fisica dell'Universit\`a di Pisa,
                       Pisa, Italy}

\vskip 0.4cm
\begin{center}
E-mail: {\tt calabres@df.unipi.it, Andrea.Pelissetto@roma1.infn.it,} \\

{\tt Ettore.Vicari@df.unipi.it}
\end{center}

\vskip 1.cm

\begin{abstract}
We discuss the critical behavior of several three-dimensional 
magnetic systems, such as  pure and randomly dilute (anti)ferromagnets and
stacked triangular antiferromagnets. 
We also discuss the nature of the multicritical points that arise in the 
presence of two distinct $O(n)$-symmetric order parameters and, in particular,
the nature of the multicritical point in the 
phase diagram of high-$T_c$ superconductors that has been predicted
by the SO(5) theory.
For each system, we consider the corresponding Landau-Ginzburg-Wilson field 
theory and review the field-theoretical results
obtained from the analysis of high-order perturbative
series in the frameworks of the $\epsilon$ and of the fixed-dimension $d=3$
expansions. 
\end{abstract}

\end{titlepage}

\section{Introduction}
\label{lsec-intro}

In the framework of the renormalization-group (RG)
approach to critical phenomena,
a quantitative description of 
many continuous phase transitions  can be obtained
by considering an effective Landau-Ginzburg-Wilson (LGW) theory,
containing up to fourth-order powers of the field components.
The simplest example is the O($N$)-symmetric $\phi^4$ theory,
\begin{equation}
{\cal H}_{O(N)} = \int d^d x \Bigl[ 
{1\over 2} \sum_i (\partial_\mu \Phi_{i})^2 + 
{1\over 2} r \sum_i \Phi_{i}^2  + 
{1\over 4!} u \sum_{ij} \Phi_i^2 \Phi_j^2 \Bigr], 
\label{HON}
\end{equation}
where $\Phi$ is an $N$-component real field. This model
describes several universality classes:
the Ising one for $N=1$ 
(relevant for the liquid-vapor transition in simple fluids, for the 
Curie transition in uniaxial magnetic systems, etc...), 
the XY one for $N=2$
(relevant for the superfluid transition in $^4$He, 
and the magnetic transition in magnets with easy-plane anisotropy, etc...),
the Heisenberg one for $N=3$
(it describes the Curie transition in isotropic magnets), 
the O(4) universality class for $N=4$
(relevant for the finite-temperature transition
in two-flavor quantum chromodynamics, the theory of strong interactions);
in the limit $N\rightarrow 0$ it describes the behavior of dilute homopolymers
in a good solvent in the limit of large polymerization.
See, e.g., Refs.~\cite{Zinn-Justin-book,review} for 
recent reviews.
But there are  also other physically interesting 
transitions described by LGW field theories
characterized by more complex symmetries.

The general LGW Hamiltonian for an $N$-component field $\Phi_i$
can be written as
\begin{equation}
{\cal H} = \int d^d x \Bigl[ 
{1\over 2} \sum_i (\partial_\mu \Phi_{i})^2 + 
{1\over 2} \sum_i r_i \Phi_{i}^2  + 
{1\over 4!} \sum_{ijkl} u_{ijkl} \; \Phi_i\Phi_j\Phi_k\Phi_l \Bigr], 
\label{generalH}
\end{equation}
where the number of independent parameters $r_i$ and $u_{ijkl}$ 
depends on the symmetry group of the theory. An interesting 
class of models are those in which $\sum_i \Phi^2_i$ is the 
only quadratic invariant polynomial.
In this case, all $r_i$ are equal, $r_i = r$, and 
$u_{ijkl}$ satisfies the trace condition~\cite{BLZ-76}
\begin{equation}
\sum_i u_{iikl} \propto \delta_{kl}.
\label{traceco}
\end{equation}
In these models, criticality is driven by tuning the single parameter
$r$. Therefore, they describe critical phenomena characterized 
by one (parity-symmetric) relevant parameter, 
which often corresponds to the temperature. Of course, there is also
(at least one) parity-odd relevant parameter that corresponds to 
a term $\sum_i h_i \Phi_i$ that can be added to the Hamiltonian
(\ref{generalH}). For symmetry reasons, criticality occurs
for $h_i\to 0$. 
There are several physical systems whose critical behavior 
can be described by this type of LGW Hamiltonians with two or more quartic 
couplings, see, e.g., Refs.~\cite{Aharony-76,review}.
More general LGW Hamiltonians, that allow for the presence
of independent quadratic parameters $r_i$, must be considered
to describe multicritical behaviors 
arising from the competition of distinct types of ordering.
A multicritical point can be observed at the intersection of 
two critical lines with different order parameters.
In this case the multicritical behavior is achieved
by tuning two relevant scaling fields, which may correspond to
the temperature and to an anisotropy parameter \cite{KNF-76}.

In the field-theoretical (FT) approach the RG flow 
is determined by a set of RG
equations for the correlation functions of the order parameter.
In the case of a continuous transition, 
the critical behavior is determined by the stable fixed point (FP)
of the theory, which characterizes a universality class. 
The absence of a stable FP may instead be considered
as an indication for a first-order transition,
even in those cases in which the mean-field approximation predicts
a continuous transition.
But, even in the presence of a stable FP,
a first-order transition may  still occur
for systems that are outside its attraction domain. 
The RG flow can be studied by perturbative methods, by performing an 
expansion in powers of $\epsilon\equiv 4-d$ \cite{WF-72} or a 
fixed-dimension (FD) expansion in powers of appropriate zero-momentum
quartic couplings \cite{Parisi-80}. It is also possible to combine 
the FD approach with the $4-d$ regularization, obtaining the 
minimal-subtraction scheme without $\epsilon$
expansion \cite{Dohm-85}.
In these perturbative approaches,
the computation and the resummation of high-order series
is essential to obtain reliable results for three-dimensional transitions
(see Refs.~\cite{Zinn-Justin-book,review} for reviews). 
Beside improving the accuracy, in some cases high-order
calculations turn out to be necessary to determine 
the correct physical picture in three dimensions.

In this paper we discuss the applications of the FT approach
to the study of several three-dimensional transitions
in magnetic systems.
Beside the O($N$)-symmetric Hamiltonian ${\cal H}_{O(N)}$ that 
is relevant for magnetic transition in uniaxial, easy-plane, and isotropic
magnets, we consider LGW Hamiltonians
characterized by more complex symmetries: in particular, 
we consider Hamiltonians invariant under the cubic group,
under O($M$)$\otimes$O($N$) and O($M$)$\oplus$O($N$).
They allow us to investigate 
the effects of the addition of uncorrelated nonmagnetic impurities
to magnetic systems, 
the critical behavior of frustrated spin models with noncollinear order---we 
can check whether a new chiral universality class exists---and the nature of 
the multicritical points arising 
from the competition of distinct $O(n)$ order parameters.
We present an overview of results obtained by using
perturbative FT methods and compare them
with other theoretical approaches that provide results of 
comparable precision---first of all, with results obtained using 
lattice techniques, e.g., Monte Carlo and high-temperature expansions---and 
with experiments.

The paper is organized as follows.
Sec.~\ref{on} is dedicated to the standard O($N$) universality classes.
We present a brief overview of FT results for the
universality classes that are relevant for magnetic systems,
such as the Ising, XY, and Heisenberg ones.
In Sec.~\ref{onstab} we analyze the stability of the 
O($N$)-symmetric FP under generic perturbations in  
three-dimensional $N$-component systems.
This analysis is useful to assess the relevance
of magnetic anisotropies in the critical behavior of physical systems.
In Sec.~\ref{random} we discuss the effects of the addition of quenched 
nonmagnetic impurities on the critical behavior of magnetic systems,
focusing on pure systems that have transitions belonging to the 
standard O($N$) universality classes.
In Sec.~\ref{chiral} we consider frustrated spin models
with noncollinear order, whose critical behavior is effectively described
by an O($M$)$\otimes$O($N$)-symmetric Hamiltonian with $M=2$.
These models are relevant for stacked triangular antiferromagnets.
In Sec.~\ref{multi} we consider
the O($n_1$)$\oplus$O($n_2$)-symmetric $\phi^4$ theory that describes 
the universal behavior near the multicritical point 
where two critical lines with symmetry O($n_1$) and O($n_2$) meet.
Finally, we discuss the relevance of these results for 
anisotropic antiferromagnets in the presence 
of a magnetic field and for high-$T_c$ superconductors
within the so-called SO(5) theory.

\section{The O($N$) universality classes}
\label{on}

The O($N$) universality classes are relevant for
magnetic systems, because they describe the critical behaviors in 
uniaxial, easy-plane, and isotropic magnets,
corresponding respectively to the Ising ($N=1$), XY ($N=2$)
and Heisenberg ($N=3$) universality classes.
The universal critical properties can be investigated by using
perturbative FT methods applied to the
O($N$)-symmetric Hamiltonian (\ref{HON}).
The oldest perturbative method is the $\epsilon$ expansion in which 
the expansion parameter is $\epsilon \equiv 4 - d$ \cite{WF-72}. 
Subsequently, 
Parisi \cite{Parisi-80} pointed out the possibility of using 
pertubation theory directly at the physical dimensions $d=3$ and $d=2$
in the massive (high-temperature) phase.

\subsection{The fixed-dimension expansion}
\label{sec-2.4.1}

In the FD expansion one works directly in $d=3$ or 
$d=2$. In this case the theory is super-renormalizable since the 
number of primitively divergent diagrams is finite. 
One may regularize the corresponding  integrals by keeping $d$ arbitrary 
and performing an expansion in $\epsilon=3-d$ or $\epsilon=2-d$. 
Poles in $\epsilon$ appear in divergent diagrams. Such divergences
are related to the necessity of performing an infinite renormalization of the 
parameter $r$ appearing in the bare Hamiltonian, see, e.g., the discussion
in Ref.~\cite{BB-85}. This problem
can be avoided by replacing the quadratic parameter $r$ of the Hamiltonian
with the mass $m$ (inverse second-moment correlation length) 
defined by 
\begin{equation}
m^{-2} = \, {1\over \Gamma^{(2)}(0)} \, 
     \left. {\partial \Gamma^{(2)}(p^2) \over \partial p^2}\right|_{p^2=0},
\end{equation}
where the function $ \Gamma^{(2)}(p^2)$ is related 
to the one-particle irreducible two-point function  
(i.e. the inverse two-point function of the order-parameter field) by
\begin{equation}
\Gamma^{(2)}_{ab} (p) = \delta_{ab} \Gamma^{(2)}(p^2).
\end{equation}
Perturbation theory in terms of $m$ and $u$ is finite in $d<4$. 
The critical limit is obtained for 
$m\to 0$. To handle it, one considers appropriate 
RG functions. Specifically, one defines
the zero-momentum four-point coupling $g$ and the
field-renormalization constant $Z_\phi$ by
\begin{eqnarray}
&&\Gamma^{(2)}_{ab}(p) = \delta_{ab} Z_\phi^{-1} \left[ m^2+p^2+O(p^4)\right],
\label{ren1} \\
&&\Gamma^{(4)}_{abcd}(0) = 
Z_\phi^{-2} m^{4-d}
{g\over 3}\left(\delta_{ab}\delta_{cd} + \delta_{ac}\delta_{bd} +
                \delta_{ad}\delta_{bc} \right),
\label{ren2}
\end{eqnarray}
where $\Gamma^{(n)}_{a_1,\ldots,a_n}$ are one-particle irreducible 
correlation functions. 
Then, one defines a coupling-renormalization constant $Z_g$  and a
mass-renormalization constant $Z_t$ by
\begin{eqnarray}
u = m^{4-d} g Z_g Z_\phi^{-2}, \qquad
\Gamma^{(1,2)}_{ab}(0)=\, \delta_{ab} Z_t^{-1}, \label{ren3}
\end{eqnarray}
where $\Gamma^{(1,2)}_{ab}(p)$ is the one-particle irreducible
two-point function with an insertion of ${1\over2}\phi^2$.
The renormalization constants are determined as perturbative expansions
in powers of $g$. The FP of the model is determined by
the nontrivial zero $g^*$ of the $\beta$-function
\begin{eqnarray}
   \beta(g) = m \left. {\partial g\over \partial m}\right|_{u} 
=   (d-4) g\left[ 1 + g {d\over dg}\log (Z_g Z_\phi^{-2})\right]^{-1}.
\end{eqnarray}
Then, one defines
\begin{eqnarray}
&&\eta_\phi(g) = \left. {\partial \ln Z_\phi \over \partial \ln m}
         \right|_{u}
= \beta(g) {\partial \ln Z_\phi \over \partial g},
\\
&&\eta_t(g) = \left. {\partial \ln Z_t \over \partial \ln m}
         \right|_{u}
= \beta(g) {\partial \ln Z_t \over \partial g}.
\end{eqnarray}
Finally, the critical exponents are given by
\begin{eqnarray}
\eta &=& \eta_\phi(g^*),
\label{eta_fromtheseries} \\
\nu &=& \left[ 2 - \eta_\phi(g^*) + \eta_t(g^*)\right] ^{-1},
\label{nu_fromtheseries} \\
\omega &=& \beta'(g^*),
\end{eqnarray}
where $\omega$ is the exponent associated with the
leading irrelevant operator.
All other exponents can be obtained using the scaling 
and hyperscaling relations.

In three dimensions the $\beta$-function is known to six loops 
\cite{BNGM-77}, while the RG functions associated with the critical exponents
have been computed to seven loops \cite{MN-91}.
FT perturbative expansions are divergent. Thus,
in order to obtain accurate results, an appropriate resummation 
is required. This can be done by exploiting their Borel summability, 
and the knowledge of the high-order behavior of the expansion,
see, e.g., Ref.~\cite{LZ-77}.
In Table~\ref{onexp} we report some recent results for the critical
exponents of the Ising, XY, and Heisenberg universality classes. 
For comparison, we also report estimates 
from lattice techniques, such as high-temperature (HT) expansions
and Monte Carlo (MC) simulations.
The agreement is very good. 
A much more complete list of results and references
can be found in Ref.~\cite{review}.
Of course, it is important to note that there is also a good
general agreement with experiments, see, e.g., the list of results
reported in Ref.~\cite{review}.

Since this FD expansion is based on zero-momentum renormalization 
conditions, it is not suited for the study of 
the low-temperature (broken) phase. In order to investigate the
low-temperature phase other perturbative FD schemes should be considered,
such as the minimal-subtraction scheme without $\epsilon$ 
expansion \cite{Dohm-85}.

\subsection{The $\epsilon$ expansion} 
\label{sec-2.4.2}

The $\epsilon$ expansion 
\cite{WF-72} is based on the observation that, for $d=4$, 
the theory is essentially Gaussian. One  considers 
the standard perturbative expansion, and then transforms it
into an expansion in powers of $\epsilon\equiv 4 - d$. 
In practice, the method works as in the FD expansion. 
One first determines the expansion of the
renormalization constants 
$Z_g$, $Z_\phi$, and $Z_t$ in powers of the coupling $g$. 
Initially, they were obtained by 
requiring the normalization conditions
(\ref{ren1}), (\ref{ren2}), and (\ref{ren3}). However, in this framework it 
is simpler to use the minimal-subtraction scheme \cite{tHV-72}. 
Once the renormalization constants are determined, one computes 
the RG functions $\beta(g)$, $\eta_\phi(g)$, and $\eta_t(g)$ as in
Sec.~\ref{sec-2.4.1}. 
The FP value $g^*$ is obtained by solving the equation $\beta(g^*) = 0$
perturbatively in $\epsilon$.
Once the expansion of $g^*$ is available,
one obtains the expansion of the exponents,  by expressing
$\eta_\phi(g^*)$ and $\eta_t(g^*)$ in powers of $\epsilon$. 
In this scheme, five-loop series for the exponents were computed in 
Refs.~\cite{CGLT-83,KNSCL-93}. 
The results of their analysis, exploiting Borel summability,
are reported in Table~\ref{onexp}.
Again the results are in good agreement with 
the other approaches.

\begin{table*}
\caption{
Three-dimensional estimates of the critical exponents
from the analysis of the fixed-dimension  (FD) and $\epsilon$ expansions,
and from lattice techniques such as high-temperature (HT)
expansions and Monte Carlo simulations.
}
\label{onexp}
\footnotesize
\begin{center}
\begin{tabular}{lllllll}
\hline
\multicolumn{1}{c}{$N$}&
\multicolumn{1}{c}{}&
\multicolumn{1}{c}{$\nu$}&
\multicolumn{1}{c}{$\eta$}&
\multicolumn{1}{c}{$\alpha$}&
\multicolumn{1}{c}{$\gamma$}&
\multicolumn{1}{c}{$\beta$}\\
\hline
1 & FD exp \cite{GZ-98}& 0.6303(8)  & 0.0335(25) & $\phantom{-}$0.1091(24)& 1.2396(13) & 0.3258(14) \\
  & $\epsilon$ exp \cite{GZ-98}& 0.6305(25) & 0.0365(50) &$\phantom{-}$0.108(7)&1.2380(50)&0.3265(15) \\
  & HT exp  \cite{CPRV-ising}  & 0.63012(16) & 0.03639(15)&$\phantom{-}$0.1096(5)&1.2373(2)&0.32653(10)\\
  & MC \cite{Hasenbusch-99}  & 0.6297(5) & 0.0362(8) & $\phantom{-}$0.1109(15) & 1.2366(15)&0.3262(4)\\
\hline
2 & FD exp    \cite{GZ-98}     & 0.6703(15) & 0.0354(25) & $-$0.011(4) & 1.3169(20) & 0.3470(11) \\
  & $\epsilon$ exp \cite{GZ-98}& 0.6680(35) & 0.0380(50) & $-$0.004(11) & 1.3110(70) & 0.3467(25) \\
  & HT exp \cite{CHPRV-01} & 0.67155(27) & 0.0380(4) & $-$0.0146(8) & 1.3177(5) & 0.3485(2) \\
  & MC \cite{CHPRV-01} & 0.6716(5) & 0.0380(5) & $-$0.0148(15) & 1.3177(10) & 0.3486(3) \\\hline
3 & FD exp \cite{GZ-98} & 0.7073(35) & 0.0355(25) & $-$0.122(10) & 1.3895(50) & 0.3662(25)\\
  & $\epsilon$ exp \cite{GZ-98} & 0.7045(55) & 0.0375(45) & $-$0.114(16) & 1.382(9) & 0.3655(50)\\
  & HT exp \cite{CHPRV-02} & 0.7112(5) & 0.0375(5) & $-$0.1336(15) & 1.3960(9) & 0.3689(3) \\
  & MC \cite{CHPRV-02} & 0.7113(11) & 0.0378(6) & $-$0.1339(33) & 1.3957(22) & 0.3691(6)\\
\hline
\end{tabular}
\end{center}
\end{table*}

\section{Stability of the O($N$)-symmetric fixed point in 
3-D multicomponent systems}
\label{onstab}

In order to discuss the stability of the O($N$) FP 
 in a generic $N$-component system, 
we consider the general problem of an
O($N$)-symmetric Hamiltonian in the
presence of a perturbation $P$, i.e.,
\begin{equation}
{\cal H} = \int d^d x \left[
\case{1}{2} ( \partial_\mu \Phi)^2  + \case{1}{2} r \Phi^2  
+ \case{1}{4!} u (\Phi^2)^2 + h_p P \right],
\end{equation}
where $\Phi$ is an $N$-component field and
$h_p$ an external field coupled to $P$.
Assuming $P$ to be an eigenoperator of the RG transformations,
the singular part of the Gibbs free energy for the reduced temperature
$t\to 0$ and  $h_p \to 0$ can be written as
\begin{equation}
{\cal F}_{\rm sing}(t,h_p) \approx |t|^{d\nu}
 \widehat{\cal F} \left(h_p |t|^{-\phi_p}\right),
\end{equation}
where $\phi_p \equiv y_p \nu$ is the crossover exponent associated with the 
perturbation $P$, $y_p$ is the RG dimension of $P$,
and $\widehat{\cal F} (x)$ is a scaling function.
If $y_p>0$ the pertubation is relevant and its presence
causes a crossover to another critical behavior or to a first-order transition.

In order to discuss the stability of the O($N$) FP  in general, 
we must consider any perturbation of the O($N$) FP. 
We shall first consider perturbations that 
are polynomials of the field $\Phi^a$. Any such perturbation can be 
written \cite{Wegner-72} as a sum of terms $P_{k,l}^{a_1,\ldots,a_{l} }$, 
$k\ge l$, which are 
homogeneous in $\Phi^a$ of degree $k$ and transform as the $l$-spin 
representation of the O($N$) group. Explicitly, we have 
\begin{equation}
    P_{k,{l}}^{a_1,\ldots,a_{l}} = (\Phi^2)^{(k-l)/2} Q_{l}^{a_1,\ldots,a_{l}}
\label{pml}
\end{equation}
where $Q_{l}^{a_1,\ldots,a_{l}}$ is a homogeneous polynomial of degree $l$ 
that is symmetric and traceless in the $l$ indices. The lowest-order 
polynomials are
\begin{eqnarray}
&& Q_{1}^a = \Phi^a,
\label{spin1} \\
&&Q^{ab}_{2} =  \Phi^a \Phi^b - {1\over N} \delta^{ab} \Phi^2
 \label{spin2}\\
&& Q_{3}^{abc} = 
  \Phi^a \Phi^b \Phi^c - {1\over N + 2} \Phi^2 
   \left(\Phi^a \delta^{bc} + \Phi^b \delta^{ac} + \Phi^c \delta^{ab}\right),
\label{spin3} \\
&&Q^{abcd}_{4} = \Phi^a \Phi^b \Phi^c \Phi^d  
\nonumber \\
&&\;  - {1\over N+4} \Phi^2 \left( 
        \delta^{ab} \Phi^c \Phi^d + \delta^{ac} \Phi^b \Phi^d + 
        \delta^{ad} \Phi^b \Phi^c + \delta^{bc} \Phi^a \Phi^d + 
        \delta^{bd} \Phi^a \Phi^c + \delta^{cd} \Phi^a \Phi^b \right) 
\nonumber \\
   &&\;  + {1\over (N+2)(N+4)} (\Phi^2)^2 \left(
         \delta^{ab} \delta^{cd} + \delta^{ac} \delta^{bd} + 
         \delta^{ad} \delta^{bc} \right).
 \label{spin4}
\end{eqnarray}
The classification in terms of spin values is particularly convenient,
since polynomials with different spin do not mix under RG transformations. 
On the other hand, operators with different $k$ but with the same $l$ do mix 
under renormalization. At least near four dimensions, we can use standard 
power counting to verify that the perturbation with indices $m,l$ mixes only 
with $P_{k',l}$, $k'\le k$. In particular, $P_{l,l}$ renormalizes
multiplicatively and is therefore a RG eigenoperator. Moreover, if 
$y_{k,l}$ is the RG dimension of the appropriately subtracted $P_{k,l}$,
one can verify that for small $\epsilon$, $y_{k,l} < 0$, for $l \ge 5$, i.e.
the only relevant operators have $l \le 4$. 
It is reasonable to assume that this property holds up to $\epsilon=1$. 
In principle, one should also consider terms with derivatives of the field, 
but again one can show that they are all irrelevant 
or redundant. 
Note that the condition that $\sum \Phi^2_i$ is the only quadratic 
invariant of the theory forbids the presence in the Hamiltonian 
of any spin-2 term $P^{ab}_{2,2}$. Analogously,
the trace condition (\ref{traceco})
forbids quartic polynomials transforming as the spin-2 
representation of the O($N$) group, i.e. the operators $P_{4,2}^{ab}$.

Let us first discuss the case of perturbations that are even under parity.
Beside the O($N$)-symmetric terms $P_{2,0}=\Phi^2$ and 
$P_{4,0}=(\Phi^2)^2$ 
there are only three other polynomial perturbations that must be 
considered, i.e., $P_{2,2}^{ab}$, $P_{4,2}^{ab}$, and 
$P_{4,4}^{abcd}$.\footnote{ 
Note that, according to the above-reported discussion, $P_{2,2}^{ab}$ and 
$P_{4,4}^{abcd}$ are RG eigenoperators, while $P_{4,2}^{ab}$ 
must be in general properly subtracted, i.e. the 
RG eigenoperator is $P_{4,2}^{ab} + z P_{2,2}^{ab}$ for a suitable value of $z$.
The determination of the mixing coefficient $z$ represents a subtle point in 
the FD expansion \cite{Parisi-80}, but is trivial
in the MS scheme in $4-\epsilon$ dimensions, in which
operators with different dimensions never mix so that $z=0$.} 
According to the above-reported general arguments, the stability properties 
of the O($N$) FP under even perturbations
can be obtained by determining the RG dimensions $y_{k,l}$ of the 
five classes of operators $P_{2,0}$, $P_{2,2}^{ab}$, 
$P_{4,0}$, $P_{4,2}^{ab}$ and
$P_{4,4}^{abcd}$. Of course, the dimensions $y_{k,l}$ do not depend on the 
specific values of the indices and thus one can consider 
any particular combination for each class of operators.
In Table~\ref{ONstab} we report FT estimates of 
the RG dimensions $y_{k,l}$
for $N=2,3,4,5$, obtained from the analysis of six-loop FD
and five-loop $\epsilon$ series 
\cite{CPV-02,CPV-02-mc,GZ-98}.\footnote{Results obtained in 
other theoretical approaches and in experiments  
can be found in Refs.~\cite{CPV-02-mc,review} and references therein.}
The RG dimensions of $P_{2,0}$ and of $P_{4,0}$ can be 
computed directly in the O($N$)-invariant theory. In
particular, $y_{2,0}=1/\nu$ and $y_{4,0}=-\omega$, where $\omega$ 
is the leading irrelevant exponent in the O($N$)-invariant theory.
The quadratic perturbations $P_{2,2}^{ab}$ are relevant for the description of 
the breaking of the O($N$) symmetry down to O($M$)$\oplus$O($N-M$).
Since $y_{2,2}>0$, they are always relevant.
The RG dimension $y_{4,2}$ is negative for any $N$, so that the corresponding
spin-2 perturbation $P_{4,2}^{ab}$ is always irrelevant.
On the other hand, the sign of $y_{4,4}$ depends on $N$:
it is negative for $N=2$ and positive for $N\ge 4$.
For $N=3$ it is marginally positive, suggesting the instability
of the O(3) FP under generic spin-4 quartic perturbations.

\begin{table*}
\caption{
Three-dimensional estimates of the RG dimensions $y_{k,l}$
from $\epsilon$ and  FD expansions,
from Ref.~\cite{GZ-98} for $y_{2,0}$ and $y_{4,0}$,
from Ref.~\cite{CPV-02} for $y_{2,2}$,
from Ref.~\cite{CPV-02-mc,CPV-00} for $y_{4,2}$ and $y_{4,4}$,
and from Ref.~\cite{DPV-03} for $y_{3,3}$.
}
\label{ONstab}
\footnotesize
\begin{center}
\begin{tabular}{lcllllllll}
\hline
\multicolumn{1}{c}{$N$}&
\multicolumn{1}{c}{}&
\multicolumn{1}{c}{$y_{2,0}=\nu^{-1}$}&
\multicolumn{1}{c}{$y_{2,2}$}&
\multicolumn{1}{c}{$y_{4,0}$}&
\multicolumn{1}{c}{$y_{4,2}$}&
\multicolumn{1}{c}{$y_{4,4}$}&
\multicolumn{1}{c}{$y_{3,3}$}\\
\hline
2 & FD        & 1.493(3) & 1.766(18)& $-$0.789(11) &              & $-$0.103(8) & 0.897(15) \\ 
&$\epsilon$ & 1.497(8) & 1.766(6) & $-$0.802(18) & $-$0.624(10) &$-$0.114(4) & 0.90(2) \\
\hline

3 & FD        & 1.414(7) & 1.80(3)  & $-$0.782(13) & &  $\phantom{-}$0.013(6) & 0.96(3) \\ 
&$\epsilon$ & 1.419(11)& 1.790(3) & $-$0.794(18) & $-$0.550(14) & $\phantom{-}$0.003(4)& 0.97(4) \\ 
\hline

4& FD          & 1.350(11)& 1.82(5)  & $-$0.774(20) & & $\phantom{-}$0.111(4)& 1.03(3)  \\ 
&$\epsilon$  & 1.357(15)& 1.813(6) & $-$0.795(30) & $-$0.493(14) & $\phantom{-}$0.105(6) & 1.04(5) \\ 
\hline

5& FD & 1.312(12) & 1.83(5) & $-$0.790(15) & &  $\phantom{-}$0.189(10) & 1.07(2) \\ 
&$\epsilon$ & 1.333(36) & 1.832(8) & $-$0.783(26) & $-$0.441(13) & $\phantom{-}$0.198(11) & 1.08(4)\\ 
\hline
\end{tabular}
\end{center}
\end{table*}

Let us now briefly consider the odd perturbations.
The RG dimension $y_{1,1}$ of $P_{1,1}^a$ is given by
$y_{1,1} = (\beta + \gamma)/\nu = 5/2-\eta/2$.
The operator $P_{3,1}^a$ is redundant,
because a Hamiltonian term containing $P_{3,1}$ 
can be always eliminated by a redefinition of the field $\Phi^a$.
The  RG dimension $y_{3,3}$ of the spin-3 perturbations
$P_{3,3}^{abc}$ has been computed to six loops in the FD expansion 
and to five loops in the $\epsilon$ expansion \cite{DPV-03}.
The analyses of the series provide the estimates reported in Table~\ref{ONstab},
which show that 
the spin-3 perturbation $P_{3,3}$ is relevant for any $N$.  

The stability of the O($N$) FP under spin-4 perturbations
can also be inferred from the RG flow of the cubic-symmetric LGW Hamiltonian
for an $N$-component field
\begin{equation}
{\cal H}_{c} =  \int d^d x\, \left\{ {1\over 2} \sum_{i=1}^{N}
      \Bigl[ (\partial_\mu \Phi_i)^2 +  r \Phi_i^2 \Bigr]  
+{1\over 4!} \Bigl[ u (\sum_i^N \Phi_i^2)^2 + v \sum_i^N \Phi_i^4 \Bigr]
\right \} .
\label{Hphi4cubic}
\end{equation}
The cubic-symmetric LGW Hamiltonian
${\cal H}_c$ is physically relevant because it
takes into account the most important source of magnetic anisotropy
in Heisenberg systems.
Indeed, the magnetic interactions in crystalline solids with cubic symmetry
like iron or nickel are usually modeled by using the 
$O(3)$-symmetric Heisenberg Hamiltonian. However, this is a 
simplified model, since other interactions are present.
Among them, the magnetic anisotropy that is induced by the lattice 
structure (the so-called crystal field) is particularly relevant 
experimentally, see, e.g., Ref.~\cite{Chikazumi-book}. In cubic-symmetric
lattices it gives rise to additional single-ion contributions, the 
simplest one being $\sum_i \vec{s}^{\ 4}_i$. 
These terms are usually not considered
when the critical behavior of cubic magnets is discussed.
However, this is strictly justified only
if these nonrotationally invariant interactions, that have the 
reduced symmetry of the lattice, are irrelevant in the RG
sense. This issue can be investigated by considering  the cubic-symmetric
Hamiltonian ${\cal H}_c$ \cite{Aharony-76}.
Its RG flow has been much studied 
using various FT methods \cite{review}. 
The O($N$) FP turns out to be unstable for $N>N_c$ with $N_c\approx 3$.
The most accurate results have been provided by analyses of 
high-order FT perturbative expansions, six-loop FD 
and five-loop $\epsilon$ series, see, e.g., Refs.~\cite{CPV-00,FHY-00,KS-95}:
they find $N_c\ltapprox 2.9$ in three dimensions and the existence of 
a stable FP characterized by a reduced cubic symmetry for $N\geq N_c$.
Therefore, in three-component magnets the isotropic
FP is unstable, and the RG trajectories flow toward
a stable cubic FP,
see, e.g., Ref.~\cite{CPV-00}.
However, differences between the Heisenberg and cubic critical exponents
are very small \cite{CPV-03}, 
for example $\nu$ differs by less than 0.1\%, which is
much smaller than the  typical experimental error for Heisenberg systems
\cite{review}.
Therefore, distinguishing the cubic and the Heisenberg universality class
should be very hard in experiments and numerical Monte Carlo simulations.

The stability properties of the O($N$)-symmetric FP in the
theory defined by ${\cal H}_c$ can be extended to any
$N$-component system in the presence of spin-4 terms.
The point is that the cubic-symmetric perturbation $\sum_i \Phi_i^4$
is a particular combination of the spin-4 operators $P_{4,4}^{abcd}$ and 
of the spin-0 term $P_{4,0}$.  
Since the RG dimension of the spin-4 perturbation does not depend
on the particular component, we can conclude that 
the O($N$)-symmetric FP is unstable under any spin-4 quartic perturbation 
for $N\ge 3$. 

\section{Randomly dilute spin models}
\label{random}

\begin{figure*}[tb]
\hspace{-1.5cm}
\centerline{\psfig{width=7truecm,angle=-90,file=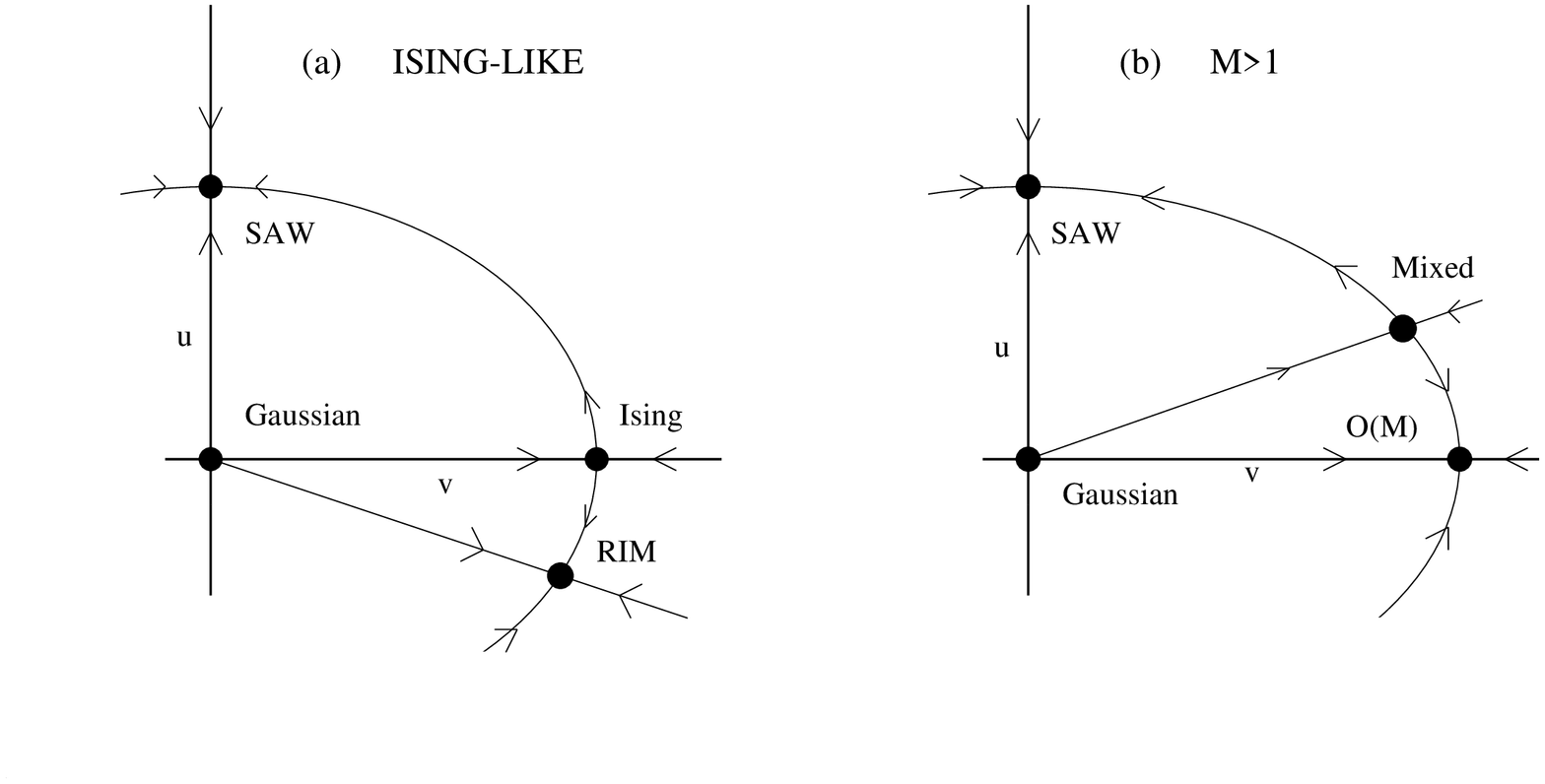}}
\vspace{-0.5cm}
\caption{
Scketch of the RG flow in the coupling plane $(u,v)$ for
(a) Ising ($M=1$) and (b) $M$-component ($M>1$) randomly dilute systems. 
}
\label{randomrgflow}
\end{figure*}

The critical behavior of systems with quenched disorder is 
of considerable theoretical and experimental interest. 
A typical example is obtained by mixing a
magnetic material with short-range interactions 
with a nonmagnetic material.
These physical system can be described by
the three-dimensional randomly dilute spin model 
\begin{equation}
{\cal H}_p = J\,\sum_{<ij>}  \rho_i \,\rho_j \; s_i \cdot s_j,
\label{latticeH}
\end{equation}
where the sum is extended over all nearest-neighbor sites,
$s_i$ are $M$-component spin variables,
and $\rho_i$ are uncorrelated quenched random variables, which are equal to one 
with probability $p$ (the spin concentration) and zero with probability $1-p$
(the impurity concentration). 
See, e.g., Refs. \cite{Aharony-76,Belanger-00,review,FHY-01} for reviews
discussing these systems.
For sufficiently low spin dilution $1-p$, i.e. above the 
percolation threshold of the spin concentration, 
the system described by the Hamiltonian ${\cal H}_p$ undergoes a second-order 
phase transition at $T_c(p) < T_c(p=1)$.

The critical behavior of randomly dilute spin systems
is rather well established. 
Randomly dilute Ising systems, of which the random Ising 
model (RIM) (\ref{latticeH}) is an example, belong 
to a new universality class which is distinct from the 
Ising universality class describing the critical behavior of the 
pure system. This has been clearly observed in experiments on dilute
uniaxial antiferromagnets such as 
Fe$_x$Zn$_{1-x}$F${}_2$ and  Mn$_x$Zn$_{1-x}$F${}_2$ materials.
The critical exponents turn out to be independent from
the impurity concentration and definitely different from those
of the pure Ising universality class,
see, e.g., their estimates reported in Refs.~\cite{Belanger-00,review,FHY-01}.
In these antiferromagnetic systems the 
presence of a uniform magnetic field causes a crossover to
a different random-field critical behavior \cite{crosstorfim}.
These experimental result are  confirmed by renormalization-group studies,
showing that the pure Ising FP 
is unstable with respect to the addition of impurities and that the 
critical behavior is controlled by a new stable RIM FP.

Unlike Ising-like systems, 
multicomponent O($M$)-symmetric spin systems 
do not change their asymptotic critical behavior
in the presence of random impurities. 
This is predicted by the Harris criterion \cite{Harris-74}
which states that the addition of impurities to a system which undergoes
a continuous transition
does not change its critical behavior if the specific-heat
critical exponent $\alpha$ of the pure system is negative.
For $M>1$ the specific-heat exponent
$\alpha_M$ of the pure system is negative, 
for example \cite{CHPRV-01,CHPRV-02}
$\alpha_2\approx -0.014$ and $\alpha_3\approx -0.13$, 
respectively for the XY and Heisenberg universality classes,
and thus disorder is irrelevant.  From the RG point of view,
the Wilson-Fisher FP of the pure 
O($M$) theory is stable under the perturbation induced by the dilution.
The presence of impurities affects only the approach
to the critical regime, leaving the asymptotic 
behavior unchanged.
This is confirmed by 
several experiments that investigated the effect of disorder 
on the $\lambda$-transition of ${}^4$He (it belongs to the 
XY universality class) \cite{YC-97,ZR-99,review} and on 
Heisenberg magnets \cite{Kaul-85,KR-94,BK-97}.

The FT approach is  based on an
effective translation-invariant LGW Hamiltonian
that is obtained by using the replica method 
\cite{Emery-75,EA-75,GL-76,AIM-76}, i.e.
\begin{equation}
{\cal H}_{MN} =  
\int d^d x 
\left\{ \sum_{ia}{1\over 2} \left[ (\partial_\mu \Phi_{ai})^2 + 
         r \Phi_{ai}^2 \right] + 
  \sum_{ijab} {1\over 4!}\left( u_0 + v_0 \delta_{ij} \right)
          \Phi^2_{ai} \Phi^2_{bj} 
\right\} \; ,
\label{Hphi4rim}
\end{equation}
where $a,b=1,...M$ and $i,j=1,...N$.
In the limit $N\rightarrow 0$
the Hamiltonian ${\cal H}_{MN}$ with 
$u_0<0$ and $v_0>0$
is expected to describe the critical properties of dilute
$M$-component spin systems. Thus, their critical behavior can be
investigated by studying the RG flow of ${\cal H}_{MN}$ 
in the limit $N\rightarrow 0$.
For generic values of $M$ and $N$, the Hamiltonian ${\cal H}_{MN}$ describes
$M$ coupled $N$-vector models 
and it is usually called $MN$ model~\cite{Aharony-76}.
Figure~\ref{randomrgflow} sketches the expected
flow diagram in the quartic-coupling plane,
for Ising ($M=1$) and multicomponent ($M>1$) systems
in the limit $N\rightarrow 0$. There are four FP's: 
the trivial Gaussian one, an O($M$)-symmetric FP,
a self-avoiding walk (SAW) FP, and a mixed FP.
The SAW FP is stable and
corresponds to the $(M \times N)$-vector theory for $N\rightarrow 0$;
but it is not of interest for the critical 
behavior of randomly dilute spin models, since it is located in the region $u>0$,
while the region relevant for
quenched disordered systems corresponds to negative values of 
the coupling $u$.
The stability of the other FP's depends on the value of $M$.
Nonperturbative arguments \cite{Sak-74,Aharony-76}
show that the stability of the O($M$) FP  
is related to the specific-heat critical exponent of the O($M$)-symmetric
theory. Indeed, ${\cal H}_{MN}$ at the O($M$)-symmetric FP can be interpreted 
as the Hamiltonian of $N$ $M$-vector systems coupled by the 
O($MN$)-symmetric term. Since this interaction is the sum of the products 
of the energy operators of the different $M$-vector models,
the crossover exponent associated with the O($MN$)-symmetric quartic 
interaction is given by
the specific-heat critical exponent $\alpha_M$ of the $M$-vector model, 
independently of $N$. 
This implies that for $M=1$ (Ising-like systems) the pure 
Ising FP is unstable since $\phi = \alpha_1 > 0$, 
while for $M>1$ the O($M$) FP is stable given that $\alpha_M<0$,
in agreement with the Harris criterion.
For $M>1$  the mixed FP is in the region of positive 
$u$ and is unstable \cite{Aharony-76}.
Therefore, the RG flow 
of the $M$-component model with $M>1$ is driven towards the pure O($M$) 
FP. Thus, the asymptotic behavior remains unchanged when impurities are
added. But their effect is not totally negligible because they
give rise to slowly-decaying scaling corrections 
proportional to $t^\Delta_r$ with $\Delta_r=-\alpha_M$,
where $\alpha_M$ is the specific-heat exponent of the
pure O($M$)-symmetric theory.
For Ising-like systems, the pure Ising FP is instead unstable, 
and the flow for negative
values of the quartic coupling $u$ leads to the stable mixed or 
random (RIM) FP which is located in the region of negative values of $u$.
The above picture emerges clearly in the framework of the $\epsilon$ expansion,
although the RIM FP is 
of order $\sqrt{\epsilon}$~\cite{Khmelnitskii-75} rather than $\epsilon$.

\begin{figure}[tb]
\centerline{\psfig{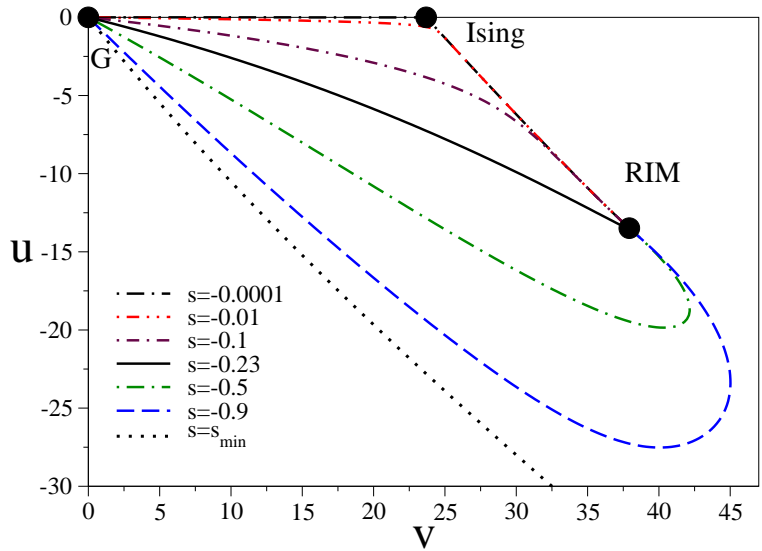}}
\vspace{2mm}
\caption{
RG trajectories in the $u,v$ plane for various values of $s\equiv u_0/v_0<0$.
From Ref.~\cite{CPPV-03}.
}
\label{trajrim}
\end{figure}

\begin{figure}[tb]
\centerline{\psfig{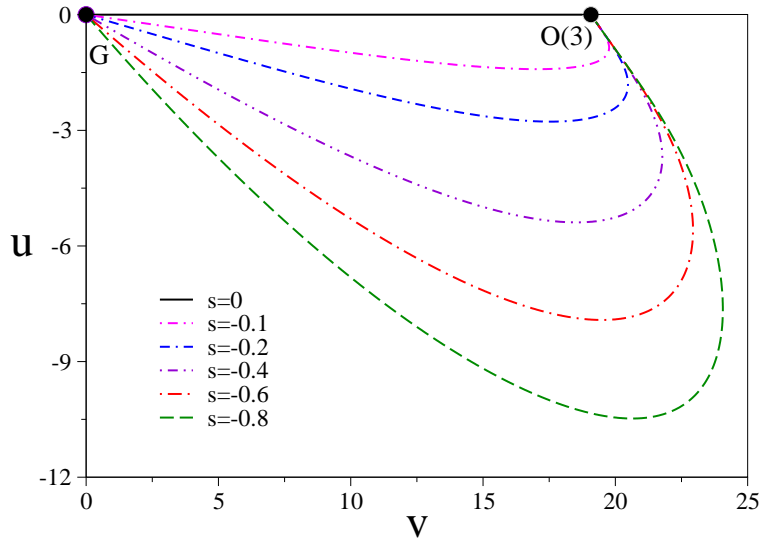}}
\vspace{2mm}
\caption{
RG trajectories of the dilute Heisenberg model
for various values of 
$s\equiv u_0/v_0<0$. From Ref.~\cite{CPPV-03}.}
\label{traj3}
\end{figure}

The most precise FT results have been obtained
in the framework of the FD
expansion  in powers of the zero-momentum quartic couplings
related to $u_0$ and $v_0$.
In this scheme the theory is renormalized by introducing
a set of zero-momentum conditions for the one-particle irreducible
two-point and four-point correlation functions,
which relate the zero-momentum mass scale $m$ 
and the quartic couplings $u$ and $v$ to the corresponding Hamiltonian parameters
$r$, $u_0$, and $v_0$.
This is just a straightforward extension of the method
outlined in Sec.~\ref{sec-2.4.1} to a theory with two quartic parameters.
The critical behavior of dilute spin systems is determined by the stable FP
of the theory, that is given by
the common  zero $u^*,v^*$ of the $\beta$-functions in the limit $N\rightarrow 0$
\begin{equation}
\beta_u(u,v) = \left. m{\partial u\over \partial m}\right|_{u_0,v_0} ,\qquad
\beta_v(u,v) = \left. m{\partial v\over \partial m}\right|_{u_0,v_0} ,
\end{equation}
whose stability matrix has positive eigenvalues
(actually a positive real part is sufficient).
Then, the critical exponents are obtained by evaluating appropriate RG
functions at $u^*,v^*$.
Figs.~\ref{trajrim} and \ref{traj3} show typical RG trajectories
in the zero-momentum quartic plane $u$, $v$ in the relevant
region corresponding to $u_0<0$, for the RIM and the random Heisenberg
model respectively \cite{CPPV-03}.

The FD expansions of the $\beta$ functions and of the critical exponents
have been computed to six loops \cite{PV-00,CPV-00,PS-00}.
The results of the six-loop analyses 
are reported in Table~\ref{rimexp}, where they are 
compared with estimates obtained in Monte Carlo simulations of the 
RIM, see, e.g., Refs.~\cite{CMPV-03,BFMMPR-98},
and in experiments on uniaxial antiferromagnets,
such as Fe$_x$Zn$_{1-x}$F${}_2$ and  Mn$_x$Zn$_{1-x}$F${}_2$, 
see, e.g., Refs. \cite{Belanger-00,FHY-01}.
The agreement is satisfactory, although we note
the slightly smaller value of the experimental estimate
of $\alpha$. We mention that
several computations have also been done in the framework
of the $\epsilon$ expansion.
The $\sqrt{\epsilon}$ expansion \cite{Khmelnitskii-75} turns out not to be 
effective for a quantitative
study of the RIM (see, e.g., the analysis of the five-loop series done in 
Ref. \cite{SAS-97}).
The related minimal-subtraction renormalization scheme without 
$\epsilon$ expansion \cite{Dohm-85} have been also considered.  
The three- and four-loop 
results turn out to be  in reasonable agreement with the estimates obtained 
by other methods, but at five loops no random FP 
is found \cite{FHY-00-r}.
Nonperturbative approaches based on scaling-field expansions and on continuous
RG equations have also been exploited \cite{NR-82,TMVD-01},
providing substantially consistent results.
A more complete list of 
theoretical and experimental results can be found 
in Refs.~\cite{review,Belanger-00,FHY-01}.

\begin{table*}
\caption{
Critical exponents for the RIM universality class.}
\label{rimexp}
\footnotesize
\begin{center}
\begin{tabular}{ccllll}
\hline
\multicolumn{1}{c}{}& 
\multicolumn{1}{c}{}& 
\multicolumn{1}{c}{$\gamma$}& 
\multicolumn{1}{c}{$\nu$}& 
\multicolumn{1}{c}{$\alpha$}&
\multicolumn{1}{c}{$\beta$}\\ 
\hline  
six-loop FD & \cite{PV-00} & 1.330(17) & 0.678(10)  & $-$0.034(30) & 0.349(5)\\
Monte Carlo & \cite{CMPV-03}  & 1.342(6) & 0.683(3)   & $-$0.049(9) & 0.3534(15) \\
Fe$_x$Zn$_{1-x}$F$_2$  & \cite{Belanger-00}  & 1.31(3) & 0.69(1) & $-$0.10(2) & 0.359(9) \\
\hline
\end{tabular}
\end{center}
\end{table*}

Using the FT approach, one can also compute the critical exponent $\phi$ 
describing the crossover from random-dilution to random-field
critical behavior in Ising systems,
and in particular the crossover observed in dilute anisotropic
antiferromagnets when an external magnetic field is applied \cite{Belanger-00}.
The crossover exponent $\phi$ is related to the RG dimensions of
the quadratic operator $\Phi_i\Phi_j$ ($i\neq j$) in the 
limit $N\rightarrow 0$ \cite{Aharony-86}.
Six-loop computations \cite{CPV-03-phi}
provide the estimate $\phi=1.42(2)$,
which is in good agreement with the available experimental estimates,
for example $\phi=1.42(3)$ for Fe$_x$Zn$_{1-x}$F$_2$ \cite{Belanger-00}.

Finally, we mention that a perturbative six-loop analysis 
of the combined effect of impurities and magnetic anisotropy
has been reported in Ref.~\cite{CPV-03}.

\section{Frustrated spin models with noncollinear order}
\label{chiral}

In physical magnets noncollinear order is due to frustration that may arise
either because of the special geometry of the lattice, or from the competition 
of different kinds of interactions \cite{Kawamura-98}.
Typical examples of systems of the first type are 
stacked triangular antiferromagnets (STA's), 
where magnetic ions are located at each site of 
a three-dimensional stacked triangular lattice.
On the basis of the structure of the ground state, 
in an $N$-component STA one expects a 
transition associated with the breakdown of 
the symmetry from O($N$) in the HT phase to O($N-2$) in the LT phase. 
The nature of the transition, and in particular the 
existence of a new chiral universality class \cite{Kawamura-88},
is still controversial. 
On this issue, there is still much 
debate, FT methods, Monte Carlo simulations, and experiments providing 
contradictory results in many cases (see, e.g., Ref.\cite{review}
for a recent review of results).
Overall, experiments on STA's favor a continuous transition
belonging to a new chiral universality class.

The determination of an effective LGW Hamiltonian describing
the critical behavior leads to
the O($M$)$\otimes$O($N$)-symmetric theory \cite{Kawamura-88}
\begin{equation}
{\cal H}_{ch}  = \int d^d x
 \Bigl\{ {1\over2}
      \sum_{a} \Bigl[ (\partial_\mu \phi_{a})^2 + r \phi_{a}^2 \Bigr]
+ {1\over 4!}u \Bigl( \sum_a \phi_a^2\Bigr)^2
+ {1\over 4!}  v
\sum_{a,b} \Bigl[ ( \phi_a \cdot \phi_b)^2 - \phi_a^2\phi_b^2\Bigr]
             \Bigr\},
\label{LGWH}
\end{equation}
where $\phi_a$, $a=1,\ldots, M$, are $N$-component vectors. 
The case $M=2$ with $v>0$ describes  
frustrated spin models with noncollinear order.
Negative values of $v$ correspond to magnets 
with sinusoidal spin structures. XY and Heisenberg systems
correspond to $N=2$ and $N=3$ respectively.
Recently the Hamiltonian (\ref{LGWH}) has also been considered
to discuss the phase diagram of Mott insulators \cite{Sachdev-02}.
See Refs.~\cite{Kawamura-98,review} for other applications.

Six-loop calculations \cite{PRV-01} in the framework of the
$d=3$ FD expansion provide a rather robust evidence
for the existence of a new stable FP in the
XY and Heisenberg cases corresponding to the conjectured
chiral universality class, contradicting earlier studies
based on much shorter (three-loop) series \cite{AS-94}.
It has also been argued \cite{CPS-02} that the stable chiral FP
is actually a focus, due to the fact that the eigenvalues of 
its stability matrix turn out to have a nonzero imaginary part.
The new chiral FP's found for $N=2,3$
should describe the apparently continuous transitions observed in STA's.
The FT estimates of the critical exponents are in satisfactory agreement with
the experimental results, including the chiral crossover exponent
related to the chiral degrees of freedom \cite{PRV-02}.

On the other hand, other FT studies, see, e.g., Refs. \cite{Zumbach-94,TDM-00},
based on approximate solutions of continuous RG equations,
do not find a stable FP, thus favoring a weak first-order transition.
Monte Carlo simulations have not been conclusive in setting the question, 
see, e.g., Refs.~\cite{LS-98,Itakura-01,PS-02}.
Since all the above approaches rely on  
different approximations and assumptions, their comparison and consistency 
is essential before considering the issue substantially understood.

Finally, we mention that the six-loop FT analysis
of the critical behavior of systems with nonplanar ordering, i.e. with $M\ge 3$,
has not find any evidence of the presence of a stable FP \cite{Parruccini-03}, 
suggesting a first-order transition. 
A high-order FT study 
of two-dimensional systems have been reported in Ref.~\cite{CP-01}.

\section{Competition of two distinct order parameters}
\label{multi}

The competition of distinct types of ordering gives rise to multicritical
behavior. More specifically, a multicritical point (MCP) is observed at the 
intersection of two critical lines characterized by different order parameters.
MCP's arise in several physical contexts.  The phase diagram of anisotropic 
antiferromagnets in  a uniform magnetic 
field $H_\parallel$ parallel to the anisotropy axis presents
two critical lines in the temperature-$H_\parallel$ plane, 
belonging to the XY and Ising universality 
classes, that meet at a MCP \cite{FN-74,KNF-76}.
MCP's are also expected 
in the temperature-doping phase diagram of
high-$T_c$ superconductors.
Within the SO(5) theory \cite{Zhang-97,ZHAHA-99}  
of high-$T_c$ superconductivity, it has been speculated that 
the antiferromagnetic and superconducting transition lines meet
at a MCP in the temperature-doping phase diagram.

\begin{figure}[tb]
\centerline{\psfig{width=7truecm,angle=0,file=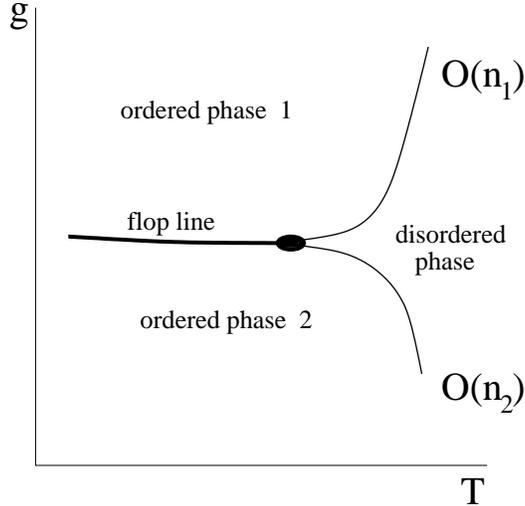}}
\vspace{2mm}
\caption{
Phase diagram in the plane $T$-$g$ presenting
a bicritical point. Here, $T$ is the temperature and $g$ 
a second relevant parameter.
The thick line (``flop line")
represents a first-order transition.
}
\label{bicr}
\end{figure}

\begin{figure*}[tb]
\centerline{\psfig{width=7truecm,angle=0,file=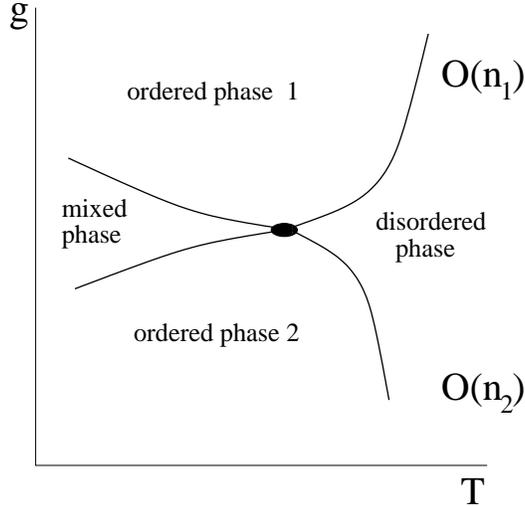}}
\caption{Phase diagram with a tetracritical point.
}
\label{tetra}
\end{figure*}

Different phase diagrams have been observed close to a MCP.
If the transition at the MCP is continuous, one may observe 
either a bicritical or a tetracritical behavior.
A bicritical behavior is characterized by the presence
of a first-order line that starts at the MCP and separates
the two different ordered low-temperature phases, see Fig.~\ref{bicr}.
In the tetracritical case, there exists a mixed low-temperature phase  
in which both types of ordering coexist and which is bounded
by two critical lines meeting at the MCP, see Fig.~\ref{tetra}.
It is also possible that the transition at the MCP is of first order.
A possible phase diagram is sketched in Fig.~\ref{tricr}. In this 
case the two first-order lines, which start at the MCP and separate
the disordered phase from the ordered phases, end in tricritical points
and then continue as critical lines.

\subsection{Multicritical behavior in O($n_1$)$\oplus$O($n_2$) theories}
\label{multion}

The multicritical behavior arising from the
competition of two types of ordering characterized by O($n$) 
symmetries is determined by the RG flow of  the most general
O($n_1$)$\oplus$O($n_2$)-symmetric
LGW Hamiltonian involving two fields $\phi_1$ and $\phi_2$
with $n_1$ and $n_2$ components respectively, i.e.
\cite{KNF-76} 
\begin{eqnarray}
{\cal H}_{mc} = &&\int d^d x \Bigl[ 
\case{1}{2} ( \partial_\mu \phi_1)^2  + \case{1}{2} (
\partial_\mu \phi_2)^2 + \case{1}{2} r_1 \phi_1^2  
 + \case{1}{2} r_2 \phi_2^2 \label{bicrHH} \\
&& 
+ u_1 (\phi_1^2)^2 + u_2 (\phi_2^2)^2 + w \phi_1^2\phi_2^2 \Bigr].
\nonumber
\end{eqnarray}
The critical behavior at the MCP is determined 
by the stable FP when both $r_1$ and $r_2$ 
are tuned to their critical value.
An interesting possibility is that the stable FP has O($N$) symmetry, 
$N\equiv n_1 + n_2$, so that the symmetry gets effectively enlarged  
approaching the MCP.
As we shall see, this is realized only in the case
$N=2$, i.e. when two Ising lines meet.

The phase diagram of the model with
Hamiltonian (\ref{bicrHH}) has been investigated
within the mean-field approximation in Ref. \cite{LF-72}.
If the transition at the MCP is continuous, one may observe 
either a bicritical or a tetracritical behavior.
But it is also possible that the transition at the MCP is of first order.
A low-order calculation in the framework of the
$\epsilon$ expansion \cite{KNF-76} 
shows that the isotropic O($N$)-symmetric FP ($N\equiv n_1 + n_2$)
is stable for $N<N_c=4-2 \epsilon + O(\epsilon^2)$.
With increasing $N$, a new FP named biconal FP (BFP),
which has only O($n_1$)$\oplus$O($n_2$) symmetry, becomes stable. 
Finally, for large $N$, the decoupled FP (DFP)
is the stable FP. In this case, the two order parameters are 
effectively uncoupled at the MCP.
The extension of these $O(\epsilon)$ results to three
dimensions suggests that 
for $n_1=1$ and $n_2=2$, the case
relevant for anisotropic antiferromagnets,
the MCP belongs to  the O(3) universality class,
while  for $n_1=2$ and  $n_2=3$,
of relevance for the SO(5) theory of high-$T_c$ superconductivity,
the stable FP is the BFP.

\begin{figure*}[tb]
\centerline{\psfig{width=7truecm,angle=0,file=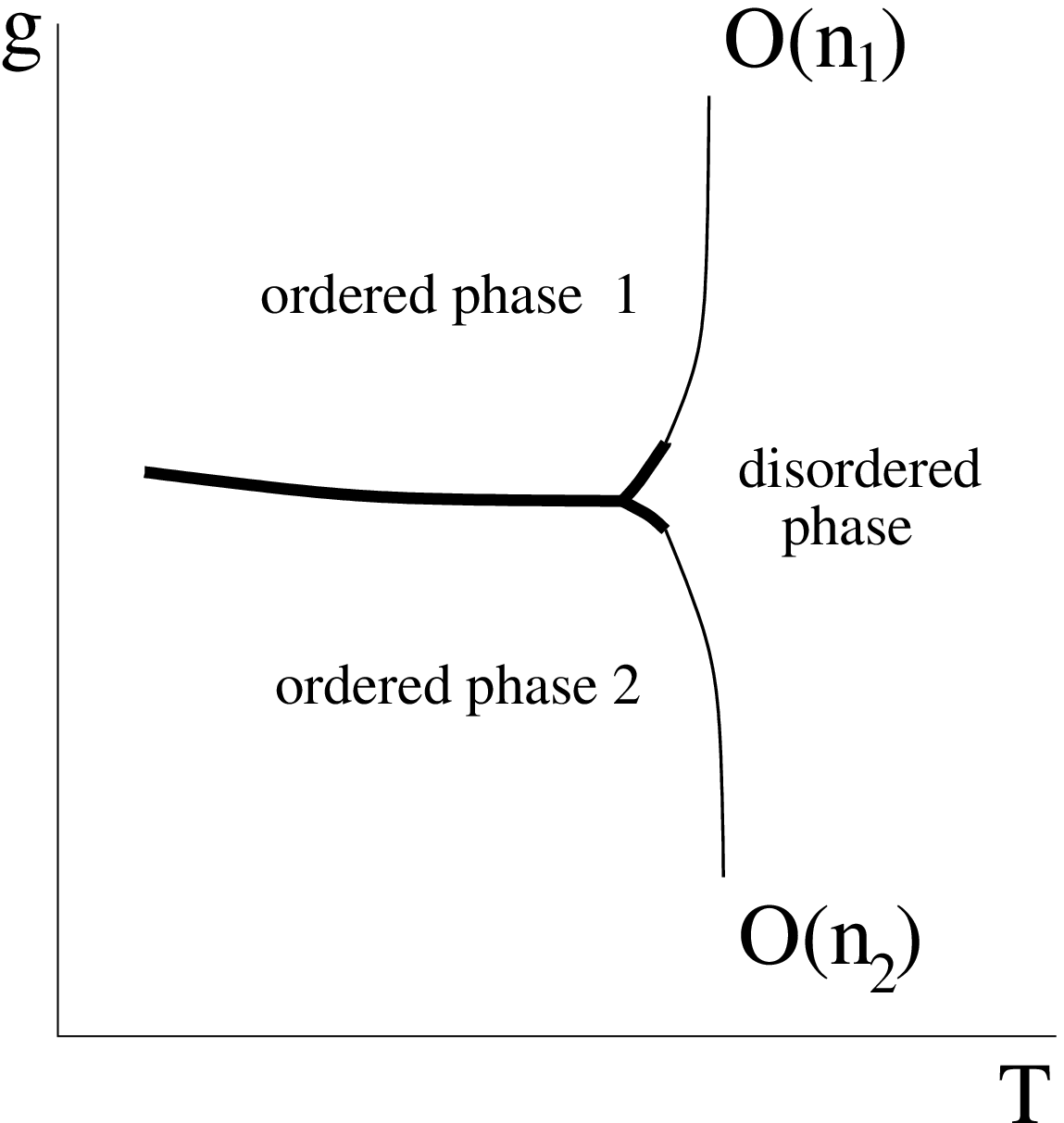}}
\caption{Phase diagram with a first-order MCP.
The thick lines represent first-order transitions.
}
\label{tricr}
\end{figure*}

The $O(\epsilon)$ computations provide useful indications on the 
RG flow in three  dimensions, but a controlled 
extrapolation to $\epsilon=1$ requires much longer series
and an accurate resummation exploiting their Borel summability.
This has been recently achieved, extending the $\epsilon$ expansion
to $O(\epsilon^5)$ \cite{CPV-02-mc}.  
A robust picture of the RG flow predicted by
the  O($n_1$)$\oplus$O($n_2$)-symmetric
LGW theory has been achieved 
by supplementing the analysis of the $\epsilon$ series 
with the results for the stability of
the O($N$) FP (cf. Sec.~\ref{onstab}),
which were also obtained by analyzing six-loop FD series,
and with nonperturbative arguments allowing to establish
the stability of the DFP \cite{Aharony-02,Aharony-02-2}.

In order to establish the stability properties of the O($N$)
symmetric FP, we first note that
the Hamiltonian (\ref{bicrHH}) contains combinations 
of spin-2 and spin-4 polynomial operators that are
invariant under the symmetry O($n_1$)$\oplus$O($n_2$).
Explicitly, they are given by
\begin{eqnarray}
&&{\cal P}_{2,0}= \Phi^2,\qquad  \qquad  
{\cal P}_{2,2}= \sum_{a=1}^{n_1} P_{2,2}^{aa} = 
                \phi_1^2-{n_1\over N} \Phi^2 , \\
&&{\cal P}_{4,0}= (\Phi^2)^2,\qquad  \qquad  
{\cal P}_{4,2}= \Phi^2 {\cal P}_{2,2},\nonumber \\
&&{\cal P}_{4,4}=  \sum_{a=1}^{n_1} \sum_{b=n_1+1}^{n_2} P_{4,4}^{aabb} = 
\phi_1^2 \phi_2^2- {\Phi^2 (n_1 \phi_2^2+n_2 \phi_1^2)\over N+4}+
{n_1 n_2 (\Phi^2)^2 \over (N+2)(N+4)},
\nonumber
\end{eqnarray}
where $\Phi$ is the $N$-component field $(\phi_1,\phi_2)$.
The RG dimension $y_{2,2}$ of ${\cal P}_{2,2}$, and therefore of the 
operator $P_{2,2}^{ab}$ (see Table~\ref{ONstab}), 
provides the crossover exponent $\phi=y_{2,2}\nu$
at the MCP.  
Concerning the fourth-order terms,
the results for the spin-4 RG dimension $y_{4,4}$ 
at the O($N$) FP
(see again Table~\ref{ONstab})
imply that the O($N$) FP is stable only for $N=2$, i.e.
when two Ising-like critical lines meet.
It is  unstable in all cases with $N\ge 3$.
This implies that for $N\ge 3$ the enlargement of the symmetry 
O($n_1$)$\oplus$O($n_2$) to O($N$) does not occur, unless
an additional parameter is tuned
beside those associated with the quadratic perturbations. 

The analysis of the five-loop $\epsilon$ series shows that
for $N=3$, i.e. for $n_1=1$ and $n_2=2$,
 the critical behavior at the MCP is described by the BFP,
whose critical exponents turn out to be very close to those of
the Heisenberg universality class.
For $N\ge 4$ and for any $n_1,n_2$
the DFP is stable, implying a tetracritical behavior.
This can also be inferred by using nonperturbative arguments 
\cite{Aharony-02,Aharony-02-2}
that allow to determine the relevant stability eigenvalue
from the critical exponents of the O($n_i$)
universality classes.
At the DFP, the dominant perturbation is the quartic coupling term 
$w\phi_1^2\phi_2^2$ that scales as
the product of two energy-like operators, which have RG dimensions
$(1-\alpha_i)/\nu_i$ where $\alpha_i$ and $\nu_i$ are the critical
exponents of the O($n_i$) universality classes. Therefore, 
the RG dimension related to the $w$-perturbation is given by
\begin{equation}
y_w = {\alpha_1\over 2\nu_1} + {\alpha_2\over 2\nu_2} = 
  {1\over \nu_1} + {1 \over \nu_2} - d.
\label{yw}
\end{equation}
The stability properties of the DFP can be established by determining
the sign of $y_w$. 
Using the estimates of the critical exponents of the three-dimension\-al
O($n_i$) universality classes, see Sec.~\ref{on},
$y_w$ turns out to be negative for $N\equiv n_1+n_2\ge 4$, 
and positive for $N=2,3$.
Therefore, the DFP turns out to be stable for $N\ge 4$.
Finally, when the initial parameters of the Hamiltonian
are not in the attraction domain
of the stable FP, the transition between the disordered
and ordered phases should be of first order in the neighborhood 
of the MCP. 
In this case, a possible phase diagram is given in Fig.~\ref{tricr},
where all transition lines are first-order close to the MCP.

\subsection{Anisotropic antiferromagnets in a uniform magnetic field}
\label{ananti}

Anisotropic antiferromagnets in a uniform magnetic field $H_\parallel$
parallel to the anisotropy axis present a MCP 
in the $T-H_\parallel$ phase diagram, where two critical lines
belonging to the XY and Ising universality classes meet \cite{KNF-76}. 
In this case the high-order FT analyses predict a multicritical 
behavior described by the BFP, in contrast with  
earlier low-order $\epsilon$-expansion calculations \cite{KNF-76}
that predicted the enlargement of the symmetry to O(3).
The mean-field approximation assigns a tetracritical behavior to the MCP
\cite{KNF-76}, but a more rigorous characterization, that requires
the computation of the corresponding scaling free energy, is needed 
to draw a definite conclusion.
Experimentally, the MCP appears to be bicritical, see, e.g., 
Refs.~\cite{RG-77,KR-79}.
A quantitative analysis of the biconal FP
shows that its critical exponents are very close to the
Heisenberg ones.
For instance, the correlation-length exponent
$\nu$ differs by less than 0.001 in the two cases.
Thus, it should be very hard to distinguish the biconal from 
the  O(3) critical behavior in experiments or numerical
works based on Monte Carlo simulations.
The crossover exponent describing the crossover from the O(3)
critical behavior 
is very small, i.e., $\phi_{4,4}=y_{4,4}\nu \approx 0.01$,
see Table~\ref{ONstab},
so that systems with a small effective breaking
of the O(3) symmetry show a very slow crossover towards
the biconal critical behavior or,
if the system is outside the attraction domain of the BFP,
towards a first-order transition; thus, they may 
show the eventual asymptotic behavior only
for very small values of the reduced temperature.

\subsection{Nature of the  multicritical point within the SO(5) theory
of high-$T_c$ superconductors.}
\label{highTc}

High-$T_c$ superconductors  are
other interesting physical systems in which MCP's may arise
from the competition of different order parameters.
At low temperatures these materials exhibit 
superconductivity and antiferromagnetism
depending on doping. The SO(5) theory \cite{Zhang-97,ZHAHA-99} 
attempts to provide a unified description of these
two phenomena, by introducing a three-component
antiferromagnetic order parameter and 
a $d$-wave superconducting order parameter with U(1) symmetry,
with an approximate O(5) symmetry.
This theory predicts 
a MCP arising from the competition of these two 
order parameters when the corresponding critical lines meet
in the temperature-doping phase diagram.
Neglecting the fluctuations of the magnetic field
and the quenched randomness introduced by doping, see,
e.g., Ref.~\cite{Aharony-02-2} for a critical discussion of this point,
one may consider  the O(3)$\oplus$O(2)-symmetric LGW Hamiltonian 
to infer the critical behavior at the MCP, see, e.g.,
Refs. \cite{HZ-00,Hu-01,BL-98,AH-00,MN-00,KAE-01}.
In particular, the analysis of Ref.~\cite{AH-00},
which uses the projected SO(5) model 
as a starting point, 
shows that one can use Eq. (\ref{bicrHH}) as an effective 
Hamiltonian.\footnote{The projected SO(5) model  \cite{ZHAHA-99} 
was introduced to overcome some inconsistencies
between the original SO(5) theory and the physics of 
the Mott insulators.}

Different scenarios have been proposed for the 
critical behavior at the MCP. 
In Refs.  \cite{Zhang-97,HZ-00,Hu-01}, it was speculated that the 
MCP is a bicritical point where the O(5) symmetry is asymptotically 
realized. This picture has been apparently supported by Monte simulations 
for a five-component O(3)$\oplus$O(2)-symmetric spin model \cite{Hu-01},
and by a quantum Monte Carlo study of the quantum projected
SO(5) model in three dimensions \cite{DAJHZ-02}.
These numerical studies show that, within the parameter
ranges considered, the scaling behavior at the MCP
is consistent with an O(5)-symmetric critical behavior.
On the basis of the $O(\epsilon)$ results 
of Refs.~\cite{KNF-76}, Refs.~\cite{BL-98,MN-00} 
predicted a tetracritical behavior governed by the
BFP.  However, since the BFP was expected to be
close to the O(5) FP, it was suggested that
at the MCP the critical exponents were in any case close to 
the O(5) ones.
These hypotheses are contradicted by the high-order FT results  
that we have discussed in Sec. \ref{multion}. They indeed
show that the DFP is the only stable FP. This implies that
the asymptotic approach to the MCP is characterized
by a decoupled tetracritical behavior or by a first-order transition
for systems that are outside the attraction domain of the DFP.
The O(5) symmetry can be asymptotically realized only by tuning 
a further relevant parameter, beside the double tuning  required to approach 
the MCP. 
The O(5) FP is unstable with a crossover exponent 
$\phi_{4,4}=y_{4,4}\nu \approx 0.15$, see Table~\ref{ONstab},
which, although rather small, is nonetheless sufficiently large  
not to exclude the possibility of observing 
the RG flow towards the eventual asymptotic behavior 
for reasonable values of the reduced temperature.
Evidence in favor of a tetracritical behavior has been recently
provided by a number of experiments, see, e.g.,
Refs.~\cite{Khay-etal-02,Lee-etal-99,Katano-etal-00,Miller-etal-02,%
Lake-etal-01,Aeppli-etal-97},
which seem to show the existence of a coexistence region 
of the antiferromagnetic and superconductivity phases.
The possible coexistence of the two phases has been discussed
in Refs.~\cite{KAE-01,ZDS-02,MOBB-00}.


\end{document}